# Robust and Fragile Entanglement of Three Qubits: Relation to Permutation Symmetry

A. K. Rajagopal and R. W. Rendell

Naval Research Laboratory, Washington D. C. 20375-5320

ABSTRACT

Entanglement properties of a basic set of eight entangled three particle pure states possessing certain permutation symmetries are studied. They fall into four sets of two entangled states, differing in their patterns of robustness to entanglement when one of the states is lost. These features are related to the permutation symmetries of the spin states of the three particles and their corresponding marginal two-particle states. It is interesting to note that the eight entangled three-qubit states discussed here are eigenstates of the three-spin Heisenberg Hamiltonian of the form $\sigma_A \cdot \sigma_B + \sigma_A \cdot \sigma_C + \sigma_B \cdot \sigma_C / 2$.

Quantum entanglement in many-particle states is one of the most striking features of quantum mechanics and is at the heart of modern quantum information processing and other intriguing phenomena arising out of this property [1]. Greenberger, Horne, and Zeilinger (GHZ) [2] announced their result concerning quantum correlations (multistate, multiparticle entanglement) which goes beyond that originally formulated by Bell and Bohm for bipartite, two particles. There have been experimental efforts to realize these correlations using photon polarizations [3] and nuclear magnetic resonance (NMR) [4]. Only in the past three years have the three-particle states been shown to have advantages over the two-particle Bell states in their application to cloning [5], teleportation [6], and dense coding [7]. It is also only recently that more detailed study of the various types of three particle entangled states have been enumerated and classified. Dür et. al. [8, 9] pointed out that the original three qubit GHZ state, while maximally entangled, is not "robust" if one of the three qubits is traced out, i.e., the remaining two-particle system is not entangled as measured by several criteria. The lack of robustness will be designated here as "fragility". They also considered another three-particle entangled state, called W, which is inequivalent to the GHZ state under stochastic local operations and classical communication. The W state is robust in that it remains entangled even after any one of the three qubits is traced out. Thus under particle loss, the entanglement properties of multi-particle systems can be quite complicated [9]. The case of mixed three particle states was examined in [10].

The purpose of this paper is to characterize the robustness of entanglement, and its relation to the permutation symmetries, for the basic set of eight entangled three particle states of spin-1/2 objects A, B, C (broadly classified as GHZ and W types). They fall into four sets of two entangled states, differing in their robustness or fragility to entanglement when one of the states is lost. These features are related to the permutation symmetries of the spin states of the three particles and their corresponding marginal two-particle states. The entanglement properties are studied using several tools for distinguishing entangled states from the nonentangled ones: the conditional Tsallis entropy criterion [11], the necessary and sufficient condition of the Peres-Horodecki criterion [12], and the concurrence [13, 14], which is related to the entanglement of formation. We focus here on the pure-state three-qubit systems.

We begin by describing the three-qubit system in terms of their classification using the composition of the three spin-1/2 states, which naturally exhibits their symmetry under permutation of the individual states. This description is useful in consideration of possible experimental verification employing an NMR set-up as in [4]. If the two states of the spin are described in terms of the polarization states of a photon as in the experiments of Bouwmeester et. al. [3], the permutation symmetry may offer an appropriate configuration of the optical arrangement of the detectors. For convenience of notation, we give the dictionary of terminology associated with these different techniques so that the results can be adopted in appropriate contexts suitably: generic - (1,0), spins - $(\uparrow,\downarrow)$, photons - (H,V), where H, V stand for horizontal and vertical polarization states of the photon beam. The generic symbol (1, 0) is often called the "computational" basis [1].

We now consider a system of three spin-1/2 objects A, B, C based on permutation symmetry to construct the canonical set of entangled three-particle states. This differs from the classification of states used in [8,10]. In the language of spins, there are 8 mutually orthonormal states classified into one quartet of 4 states with total spin 3/2 and two sets of doublet states 1, 2 each with total spin 1/2 as follows [15]:

$$|Q_1^+\rangle = |\uparrow^A \uparrow^B \uparrow^C\rangle; \quad |Q_1^-\rangle = |\downarrow^A \downarrow^B \downarrow^C\rangle;$$
$$|Q_2^+\rangle = \frac{1}{\sqrt{3}}\left\{|\uparrow^A \uparrow^B \downarrow^C\rangle + |\uparrow^A \downarrow^B \uparrow^C\rangle + |\downarrow^A \uparrow^B \uparrow^C\rangle\right\} \quad (1)$$
$$|Q_2^-\rangle = \frac{1}{\sqrt{3}}\left\{|\downarrow^A \downarrow^B \uparrow^C\rangle + |\downarrow^A \uparrow^B \downarrow^C\rangle + |\uparrow^A \downarrow^B \downarrow^C\rangle\right\}$$

The quartet states are symmetric in the permutation of any pair of particles. The division of the two doublet states is arbitrary and is done here in such a way as to make doublet 1 symmetric (S) under the permutation of the spin states of particles B and C, and doublet 2 anti-symmetric (AS) in B and C, in the same manner as in [15].

$$|D_1^+\rangle = \frac{1}{\sqrt{6}}\left\{|\uparrow^A\uparrow^B\downarrow^C\rangle + |\uparrow^A\downarrow^B\uparrow^C\rangle - 2|\downarrow^A\uparrow^B\uparrow^C\rangle\right\}$$
$$|D_1^-\rangle = \frac{1}{\sqrt{6}}\left\{|\downarrow^A\downarrow^B\uparrow^C\rangle + |\downarrow^A\uparrow^B\downarrow^C\rangle - 2|\uparrow^A\downarrow^B\downarrow^C\rangle\right\}$$
(2)

$$|D_2^+\rangle = \frac{1}{\sqrt{2}}\left\{|\uparrow^A\uparrow^B\downarrow^C\rangle - |\uparrow^A\downarrow^B\uparrow^C\rangle\right\} \quad |D_2^-\rangle = \frac{1}{\sqrt{2}}\left\{|\downarrow^A\downarrow^B\uparrow^C\rangle - |\downarrow^A\uparrow^B\downarrow^C\rangle\right\}$$
(3)

The eight states of the three particle system are now arranged according to the permutation symmetry as follows. Observing that the two quartet states $Q_1^\pm$ and the two states of the doublet 2 of eq.(3) involve only two states, we form the following mutually orthonormal combinations:

$$|GHZ^\pm\rangle = \frac{1}{\sqrt{2}}\left\{|Q_1^+\rangle \pm |Q_1^-\rangle\right\} \quad |GFR^\pm\rangle = |D_2^\pm\rangle.$$
(4)

In Eq.(4), the states divide into two subclasses according to their permutation symmetries. The first two states, GHZ$^\pm$, are symmetric under the permutation of the spin states of all three particles A, B, C among themselves, whereas the other two states, denoted here as GFR$^\pm$, are anti-symmetric with respect to permutation of the spin states of B, C. However, these tripartite states also divide into the same two subclasses according to the fragility (F) or robustness (R) of entanglement on the loss of one state as will be shown subsequently. In this paper, we focus on the relation between these robustness and symmetry properties. Our use of the nomenclature GHZ$^\pm$ is due to the similarity to the original GHZ state introduced in [2], but they could just as well have been designated GFF$^\pm$ because they are found to be fragile on the loss of any one of the states. The fragility or robustness of GFR$^\pm$, on the other hand, is found to depend on which of the three states is traced out. Generally, the eight states show different patterns of fragility and robustness and we use combinations of F and R to label these states.

Similarly, the rest of the four mutually orthonormal states are:

$$|WRR^\pm\rangle = |Q_2^\pm\rangle; \quad |WRr^\pm\rangle = |D_1^\pm\rangle. \tag{5}$$

The nomenclature W for these are also in anticipation of their common properties of the original W state [8]. Here again we have two subclasses. The first two W states, denoted here as WRR$^\pm$, are symmetric under the permutation of all three particles among themselves, just as with the first two GHZ states. These are shown subsequently to be maximally robust and the RR label indicates this. The other two, denoted as WRr$^\pm$, are symmetric under the permutation of the spin states of B and C. These are shown to have different degrees of robustness and the Rr reflects this. Fig.1 shows the state space of three qubits giving a geometrical representation of both classes of states and their permutation symmetries axes. The symmetries of the states then become transparent from the geometry and it will be seen that these symmetries are related to the robustness of the state's entanglement.

We also observe that the three-qubit states discussed here are eigenstates of the 3-spin Heisenberg Hamiltonian of the form $\sigma_A \cdot \sigma_B + \sigma_A \cdot \sigma_C + \sigma_B \cdot \sigma_C / 2$, with $(|GHZ^\pm\rangle, |WRR^\pm\rangle)$, $|GFR^\pm\rangle$, and $|WRr^\pm\rangle$ belonging respectively to the eigenvalues 5/2, -3/2, and -7/2. This observation may be of interest in constructing laboratory models of these states for purposes of experimental investigations of entangled systems [16].

We proceed to show that the pure state density matrices of all these sets of states are entangled, but that their robustness varies under the loss of one of their states according to their permutation symmetry. We employ three criteria to determine the entanglement. One is based on the negative value of the Tsallis conditional entropy [11], valid for any number of particles, and which is sufficient but not necessary for determining the state of entanglement. The second is a necessary and sufficient condition for a two qubit system or a composite of one qubit and a three-state system, based on the existence of a negative eigenvalue of its partially transposed density matrix, due to Peres and Horodecki et. al. [12]. As a third criterion, we use the concurrence [13,14], a positive value of which is a necessary and sufficient condition for the entanglement of formation. This is believed to be compatible with the Peres-Horodecki criterion [13].

We outline the calculations used in the classification of these states. We consider a pure state density matrix of three particles $\hat{\rho}_{ABC}(\phi) = |\phi(ABC)\rangle\langle\phi(ABC)|$. Their two-particle and one-particle marginal density matrices are then deduced:

$$\hat{\rho}_{AB}(\phi) = Tr_C \hat{\rho}_{ABC}(\phi), \quad \hat{\rho}_A(\phi) = Tr_{B,C} \hat{\rho}_{ABC}(\phi) \quad (A, B, C \text{ cyclic}). \quad (6)$$

From these the following three quantities are calculated to enumerate the entanglement properties of the pure state density matrices:

(i) A sufficient but not necessary condition for entanglement is the negative value of the Tsallis conditional entropy [11]

$$S_q(ABC|AB) = \frac{S_q(ABC) - S_q(AB)}{1 + (1-q)S_q(AB)}, \quad S_q(AB|A) = \frac{S_q(AB) - S_q(A)}{1 + (1-q)S_q(A)}. \quad (7)$$

and the cyclic combinations of A, B, C. Here the Tsallis entropy is given by $S_q(\hat{\rho}) \equiv Tr\{\hat{\rho}^q - \hat{\rho}\}/(1-q), q > 0.$ Here and elsewhere the Tsallis parameter q is a positive parameter which is a measure of nonadditivity. For q=1, the above results reduce to the known expressions associated with the von Neumann entropy.

(ii) The Peres-Horodecki condition [12] is necessary and sufficient for the separability of the two-qubit density matrices defined by eq.(6). It states that if one of the eigenvalues of the partial transpose of $\hat{\rho}(AB)$ is negative, then the system (AB) is entangled. Here $\hat{\rho}(AB)$ is $\rho_{mm';nn'}$ in the computational basis ($|00\rangle, |01\rangle, |10\rangle, |11\rangle$) and the partial transpose with respect to the B-states is $\left(\rho_{AB}^{T_B}\right)_{mm';nn'} = \rho_{mn';nm'}$. $\hat{\rho}(AB)$ is separable if and only if $\hat{\rho}(AB) = \sum_i p_i \hat{\rho}_i(A) \otimes \hat{\rho}_i(B)$, $0 \leq p_i \leq 1$, and $\sum_i p_i = 1$.

(iii) The concurrence C(AB) of a density matrix $\hat{\rho}(AB)$ is defined as follows [13,14]. Construct the matrix, $\hat{\tilde{\rho}}(AB) = (\hat{\sigma}_y \otimes \hat{\sigma}_y) \rho^*(AB)(\hat{\sigma}_y \otimes \hat{\sigma}_y)$ where $\hat{\rho}(AB)$ is in the standard basis as in (ii) above, $\hat{\sigma}_y$ is the standard y-component Pauli matrix, and $\hat{\rho}^*(AB)$ is the complex conjugate matrix of $\hat{\rho}(AB)$. The concurrence of $\hat{\rho}(AB)$ is then given by $C(AB) = \max\{\lambda_1 - \lambda_2 - \lambda_3 - \lambda_4, 0\}$ where $\{\lambda_1, \lambda_2, \lambda_3, \lambda_4\}$ are the square roots of the eigenvalues of the matrix product $\hat{\rho}(AB)\hat{\tilde{\rho}}(AB)$ arranged in decreasing order. The state is entangled if and only if the concurrence is positive. In addition to the two-way entanglement shared by a pair of qubits given by the concurrence, an essentially three-way entanglement shared by all three can be defined by the residual entanglement or 3-tangle, $\tau(A,B,C) = 2\left(\lambda_1^{AB}\lambda_2^{AB} + \lambda_1^{AC}\lambda_2^{AC}\right)$, where $\lambda_1^{AB}$ and $\lambda_2^{AB}$ are the square roots of the two eigenvalues of $\hat{\rho}(AB)\hat{\tilde{\rho}}(AB)$, and $\lambda_1^{AC}$ and $\lambda_2^{AC}$ are defined similarly [14].

The calculations of the three properties (i)-(iii) listed above are carried out efficiently by choosing the following parametric forms of states:

$$|\phi^I(ABC)\rangle = \alpha_1|110\rangle + \beta_1|101\rangle + \gamma_1|011\rangle, \quad |\alpha_1|^2 + |\beta_1|^2 + |\gamma_1|^2 = 1 \tag{8}$$

We obtain the following states by suitable choice of the parameters,

$$|WRR^+\rangle: \alpha_1 = \beta_1 = \gamma_1 = 1/\sqrt{3}; \quad |WRr^+\rangle: \alpha_1 = \beta_1 = 1/\sqrt{6}, \gamma_1 = -2/\sqrt{6};$$

$$|GFR^+\rangle: \alpha_1 = -\beta_1 = 1/\sqrt{2}, \gamma_1 = 0.$$

For these three states, the concurrences for the marginal pairs are $C(AB) = 2|\gamma_1||\beta_1|$, $C(AC) = 2|\gamma_1||\alpha_1|$, and $C(BC) = 2|\alpha_1||\beta_1|$. The potentially negative eigenvalues for evaluating the Peres-Horodecki criterion are $2\lambda_{AB}^{(-)} = |\alpha_1|^2 - \sqrt{|\alpha_1|^4 + C(AB)^2}$,

$2\lambda_{AC}^{(-)} = |\beta_1|^2 - \sqrt{|\beta_1|^4 + C(AC)^2}$, and $2\lambda_{BC}^{(-)} = |\gamma_1|^2 - \sqrt{|\gamma_1|^4 + C(BC)^2}$.

The next three states are:

$$|\phi^{II}(ABC)\rangle = \alpha_2|110\rangle + \beta_2|101\rangle + \gamma_2|011\rangle, \quad |\alpha_2|^2 + |\beta_2|^2 + |\gamma_2|^2 = 1 \tag{9}$$

The corresponding states are obtained by suitable choice of the parameters.

$$|WRR^-\rangle: \alpha_2 = \beta_2 = \gamma_2 = 1/\sqrt{3}; \quad |WRr^-\rangle: \alpha_2 = \beta_2 = 1/\sqrt{6}, \gamma_2 = -2/\sqrt{6};$$

$$|GFR^-\rangle: \alpha_2 = -\beta_2 = 1/\sqrt{2}, \gamma_2 = 0.$$

For these three states, the concurrences for the marginal pairs are $C(AB) = 2|\gamma_2||\beta_2|$, $C(AC) = 2|\gamma_2||\alpha_2|$, and $C(BC) = 2|\alpha_2||\beta_2|$. The potentially negative eigenvalues for evaluating the Peres-Horodecki criterion are $2\lambda_{AB}^{(-)} = |\alpha_2|^2 - \sqrt{|\alpha_2|^4 + C(AB)^2}$,

$2\lambda_{AC}^{(-)} = |\beta_2|^2 - \sqrt{|\beta_2|^4 + C(AC)^2}$, and $2\lambda_{BC}^{(-)} = |\gamma_2|^2 - \sqrt{|\gamma_2|^4 + C(BC)^2}$.

The other two states are dealt with by the parametric form

$$|\phi^{III}(ABC)\rangle = \alpha_3|111\rangle + \beta_3|000\rangle, \quad |\alpha_3|^2 + |\beta_3|^2 = 1. \tag{10}$$

Then $|GHZ^+\rangle$: $\alpha_3 = \beta_3 = 1/\sqrt{2}$ and $|GHZ^-\rangle$: $\alpha_3 = -\beta_3 = 1/\sqrt{2}$.

In this case, if the Peres-Horodecki criterion is applied to the two particle marginal density matrices, it gives only positive eigenvalues and the concurrences are found to be all zero. However, the 3-tangle is found to be τ(A, B, C) = 1. Hence the $|GHZ^\pm\rangle$ states (which, in terms of F and R, would be called $|GFF^\pm\rangle$) are fragile to the loss of any one of the qubits.

It is instructive to exhibit the eigenvalues of the various density matrices because they clearly exhibit the differences among them, from which the entanglement properties follow using criteria (i), (ii) and (iii):

$$\hat{\rho}_{ABC}(\phi^i): (1,0,0,0,0,0,0,0);$$
$$\hat{\rho}_{AB}(\phi^i): (|\alpha_j|^2, 1-|\alpha_j|^2, 0, 0)\ AC \to \beta_j; BC \to \gamma_j;$$
$$\hat{\rho}_{A}(\phi^i): (|\gamma_j|^2, 1-|\gamma_j|^2)\ B \to \beta_j; C \to \alpha_j.$$
(11)

$$\rho_{AB}^{T_{2B}}(\phi^i): (|\beta_j|^2, |\gamma_j|^2, \lambda^{(+)}, \lambda^{(-)})$$
$$2\lambda^{(\pm)} = \left(|\alpha_j|^2 \pm \sqrt{|\alpha_j|^4 + 4|\beta_j|^2|\gamma_j|^2}\right)$$
(12)
$$\rho_{AC}^{T_{2C}} \to \text{change } \beta_j \text{ to } \alpha_j; \rho_{BC}^{T_{2C}} \to \text{change } \gamma_j \text{ to } \beta_j \text{ in } \rho_{AC}^{T_{2C}}.$$

$$\hat{\rho}_{AB}(\phi^i)\hat{\tilde{\rho}}_{AB}(\phi^i): (4|\beta_j|^2|\gamma_j|^2, 0, 0, 0)$$
$$AC \to \text{change } \beta_j \text{ to } \alpha_j; BC \to \text{change } \gamma_j \text{ to } \beta_j \text{ in AC}.$$
(13)

In the above expressions, i = I, j = 1 and i = II, j = 2. The determinations of fragility and robustness patterns based on these are given in Table I.

From eq.(13), we note that all the eight pure state density matrices considered are entangled by the conditional Tsallis entropy criterion [11]. At the level of marginal two-particle states, the discussion of their entanglement is more complex, because the Tsallis conditional entropy condition evaluated from eq.(11) and the expressions given in eq.(7) are only sufficient but not necessary. We therefore also employ the Peres-Horodecki criterion and the concurrence as additional entanglement criteria. Table I gives the results for the permutation symmetries, entanglement, and robustness of all the three-state and two-state marginal systems. The $GHZ^\pm$ (i.e. $GFF^\pm$) states are found to have of all of the entanglement contained in the 3-tangle and are thus fragile to the loss of any one state. This is reflected by the zero concurrences for all of the pair states. The $GFR^\pm$ states are fragile to loss of B or C but robust to the loss of A. The $WRr^\pm$ states are robust to loss of any of the states but with reduced entanglement for loss of A, thus justifying the use of the lower case r. Finally, the $WRR^\pm$ states are robust to loss of any one state.

We now relate the various values of the concurrence to the permutation symmetries, thus establishing the relation of robustness with symmetry. It is known that in a system of N qubits, the maximum degree of entanglement that can be shared by any pair has a concurrence of 2/N and that this maximum can be attained for pure states with full permutation symmetry [17]. For the tripartite case here, this bound is 2/3 and is achieved by the fully symmetric $WRR^\pm$ states. This state symmetry is depicted in the state space diagram of Fig.1(b) and is also reflected in the full ABC symmetry which corresponds to an equilateral concurrence triangle in the entanglement space diagram of Fig. 2 (b), where the sides are proportional to the concurrences. The $WRr^\pm$ has permutation symmetry only with respect to BC and this is represented by as an isosceles triangle in Fig.1(b). These states are also robust with respect to the loss of any state, but with reduced robustness for pair BC corresponding to the less than maximal concurrence of 1/3 as shown in Fig.2(b). The $GHZ^\pm$ has full ABC symmetry as seen in Fig.1(a), yet it is completely fragile because all of the entanglement occurs in the 3-tangle and is not shared by any of the pairs. Thus C = 0 for all sides of the triangle. The concurrence triangle is thus shown as being collapsed to a point in Fig.2(a). In contrast, GFR$^{\pm}$ has the more limited permutation symmetry with respect to BC shown in Fig.1(a) with C = 0 for the sides AB and AC exhibiting fragility for these two states, whereas C = 1 for BC. Thus the concurrence triangle is just the line BC as shown in Fig.2(a). The eight tripartite states studied here, eqs.(4) and (5), fell into two classes of four GHZ-type and four W-type states according to their permutation symmetries. These permutation symmetries directly reflect the entanglement patterns of fragility and robustness. The marginals of the density matrices formed from eqs.(4) and (5) can also be seen to correspond to the maximally separable density matrix decomposition of Lewenstein et al [18], $\rho = S_{max} \rho_{sep} + (1 - S_{max}) \rho_{ent}$, in terms of a separable component with maximum separability coefficient, $S_{max}$, and a pure entangled component. The observation [19] that the concurrence is constrained by $0 < S_{max} + C \leq 1$ for bipartite states is also found for our set of basic tripartite states and this also results from their permutation symmetries. For example, the density matrices $\rho^I$ formed from $|\phi^I(ABC)\rangle$ in eq.(8) leads to the marginal $\rho_{BC}^I = Tr_A(\rho^I) = S_{max} |11\rangle\langle 11| + (1 - S_{max}) |\Psi_{BC}\rangle\langle\Psi_{BC}|$, where $|\Psi_{BC}\rangle = (\alpha_1 |10\rangle + \beta_1 |01\rangle)/\sqrt{1 - |\gamma_1|^2}$, and $S_{max} = |\gamma_1|^2$. In particular for the WRR$^+$ state, $S_{max}$ = 1/3 and $|\Psi_{BC}\rangle$ is a Bell state, and the decompositional form for $\rho_{BC}^I$ can be seen directly from the symmetry of Fig.1(b).

In conclusion, we have constructed a canonical set of three particle entangled states which fall into four classes of two states each, and each of the sets differ in their fragility or robustness under the loss of one of the states, depending on the permutation symmetry of the state. These states fall into four classes of two states each, and each of the sets differ in their fragility or robustness under the loss of one of the states, depending on the permutation symmetry of the state. The GHZ$^\pm$ and WRR$^\pm$ belong to the quartets (eq.(1)), and hence cannot be transformed into each other (a different reason is given for this feature in [8]). The GFR$^\pm$ belong to doublets 2, eq.(3), and WRr$^\pm$ belong to doublet 1, eq.(4), of the three particle system. We employed three entanglement criteria (the Tsallis conditional entropy, the Peres-Horodecki criterion and the concurrence) to determine the patterns of fragility and robustness. These patterns were shown to be related to the permutation symmetries of the corresponding states. We also observed that the three-qubit states discussed here are eigenstates of the 3-spin Heisenberg Hamiltonian of the form $\sigma_A \cdot \sigma_B + \sigma_A \cdot \sigma_C + \sigma_B \cdot \sigma_C / 2$. It is worth noting that the WRr$^\pm$ states have been investigated for use in quantum cloning [5], teleportation [6] and GHZ states for dense cloning [7].

**Acknowledgments** Both of us are supported in part by the Office of Naval Research. We also thank Dr. Peter Reynolds of ONR for supporting part of this research.

TABLE I: Classifying Three Particle (A, B, C) Entangled States by
1. Permutation symmetry (S, AS, NS), 2. Entanglement (concurrence
[AB, AC, BC] or 3-tangle [ABC]), 3. Robustness or Fragility (R, r, or F).

| States | AB | | | AC | | | BC | | | ABC | |
|---|---|---|---|---|---|---|---|---|---|---|---|
| Class | 1 | 2 | 3 | 1 | 2 | 3 | 1 | 2 | 3 | 1 | 2 |
| $GFF^\pm$ $\equiv GHZ^\pm$ | S A,B,C | 0 | F | S A,B,C | 0 | F | S A,B,C | 0 | F | S A,B,C | 1 |
| $GFR^\pm$ | NS | 0 | F | NS | 0 | F | NS | 1 | R | AS B,C | 0 |
| $WRr^\pm$ | NS | 2/3 | R | NS | 2/3 | R | S B,C | 1/3 | r | S B,C | 0 |
| $WRR^\pm$ | S A,B,C | 2/3 | R | S A,B,C | 2/3 | R | S A,B,C | 2/3 | R | S A,B,C | 0 |

# References

[1] M. A. Nielsen and I. L. Chuang, *Quantum Computation and Quantum Information,* Cambridge Univ. Press, Cambridge, (2000).

[2] D. M. Greenberger, M. Horne, and A. Zeilinger, in *Bell's theorem, Quantum Theory, and Conceptions of the Universe*, ed. M. Kafatos, Kluwer, Dordrecht 69 (1989); D. M. Greenberger, M. Horne, A. Shimony, and A. Zeilinger, Am J. Phys. **58**, 1131 (1990); N. D. Mermin, Ibid, **58**, 731 (1990).

[3] D. Bouwmeester, J. W. Pan, M. Daniell, H. Weinfurter, and A. Zeilinger, Phys. Rev. Lett. **82**, 1345 (1999).

[4] S. Lloyd, Phys. Rev. **A57**, R1473 (1998); R. J. Nelson, D. G. Cory, and S. Lloyd, Los Alamos e-print:quant-ph/9905028 (1999).

[5] D. Bruß, D. P. Vincenzo, A. Ekert, C. A. Fuchs, C. Macchiavello, and J. A. Smolin, Phys. Rev. **A57**, 2368 (1998).

[6] A. Karlsson and M. Bourennane, Phys. Rev. **A58**, 4394 (1998).

[7] J. C. Hao, C. F. Li, and G. C. Guo, Phys. Rev. **A63**, 054301 (2001).

[8] W. Dür, G. Vidal, and J. I. Cirac, Phys. Rev. **A62**, 062314 (1999).

[9] W. Dür, Phys. Rev. **A63**, 020303, (2001).

[10] A. Acin, D. Bruß, M. Lewenstein, and A. Sanpera, Classification of mixed three-qubit states, e-print: quant-ph/0103025

[11] S. Abe and A. K. Rajagopal, Physica **A289**, 157 (2001).

[12] A. Peres, Phys. Rev. Lett. **77**, 1413 (1996); M. Horodecki, P. Horodecki, and R. Horodecki, Phys. Lett. **A223**, 1 (1996).

[13] W.K. Wooters, Phys. Rev. Lett. **80**, 2245 (1998); S. Hill and W.K. Wooters, Phys. Rev. Lett. **78**, 5022 (1997).

[14] V. Coffman, J. Kundu, and W. K. Wootters, Phys. Rev. **A61**, 052306 (2000)

[15] L. I. Schiff, *Quantum Mechanics,* McGraw-Hill Book Co., Inc. New York (1949).

[16] R.G. Unanyan, B.W. Shore, and K. Bergmann, Phys. Rev. **A63**, 043405 (2001).

[17] M. Koashi, V. Buzek, and N. Imoto, Phys. Rev. **A62**, 050302 (2000).

[18] M. Lewenstein and A. Sanpera, Phys. Rev. Lett. **80**, 2261 (1998).

[19] B.-G. Englert and N. Metwally, Appl. Phys. **B72**, 35 (2001).

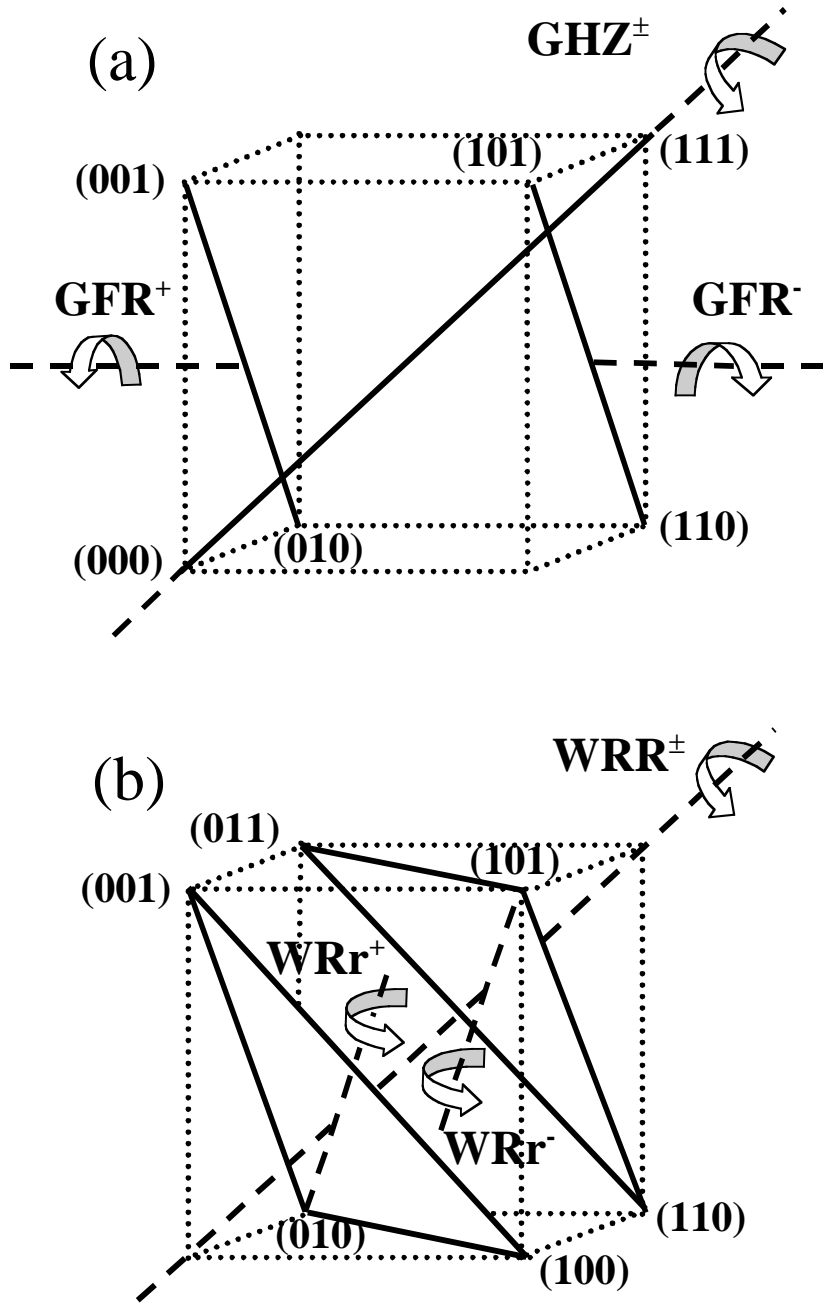

Figure 1. State space of three qubits showing geometrical representation of the eight states (solid lines) and their permutation symmetry axes (dashed lines).

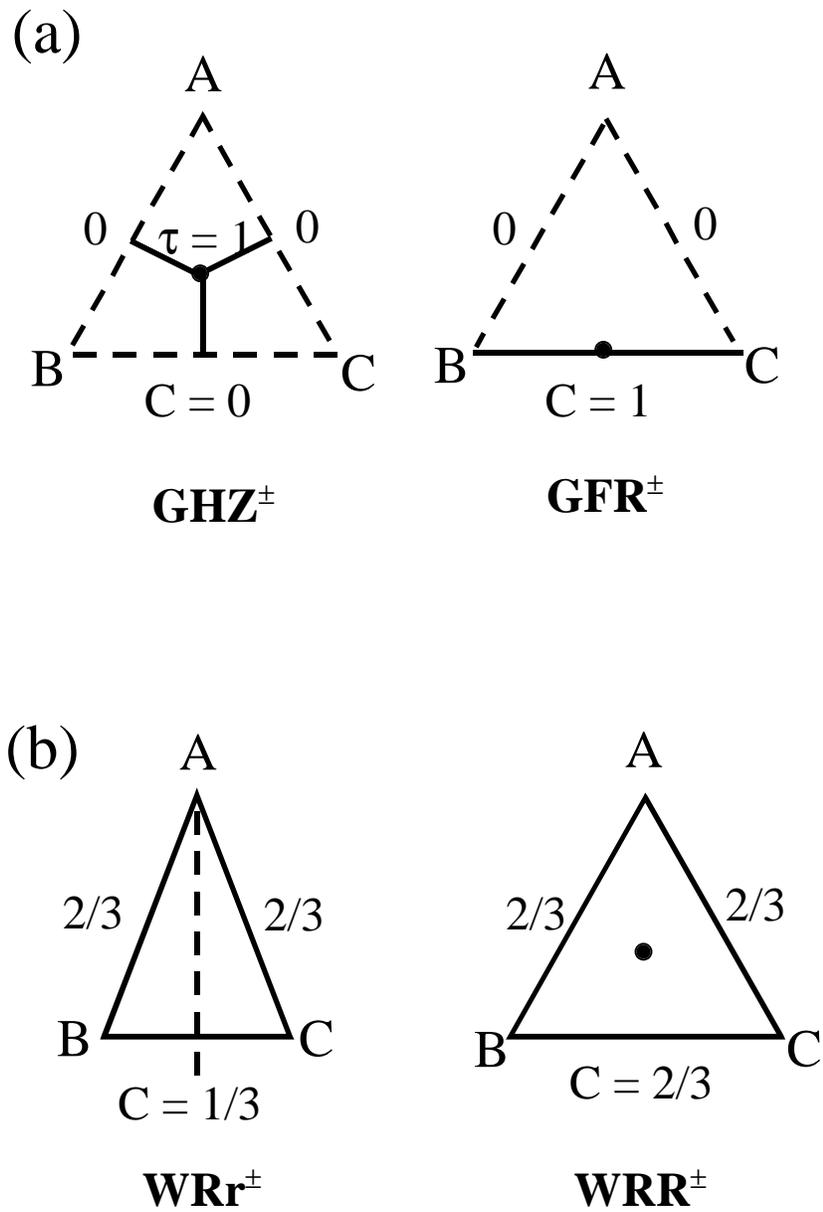

Figure 2. Entanglement space of three qubits showing the concurrence triangles corresponding to the states in Figure 1. State permutation symmetries lead to the entanglement and robustness patterns shown here and in Table I.